\documentclass[aip, amsmath,amssymb, reprint]{revtex4-1}

\usepackage{amsmath}
\usepackage{amssymb}
\usepackage[utf8]{inputenc}
\usepackage[T1]{fontenc}
\usepackage[separate-uncertainty=true]{siunitx}
\usepackage{float}
\usepackage{graphicx}

\begin{document}

\author{A.~Iorio}
\email{andrea.iorio@sns.it}
\affiliation{NEST, Istituto Nanoscienze-CNR and Scuola Normale Superiore, I-56127 Pisa, Italy}
\author{E.~Strambini}
\affiliation{NEST, Istituto Nanoscienze-CNR and Scuola Normale Superiore, I-56127 Pisa, Italy}
\author{G.~Haack}
\affiliation{Department of Applied Physics, University of Geneva, Chemin de Pinchat 22, 1227 Carouge, Gen\`eve, Switzerland}
\author{M.~Campisi}
\affiliation{NEST, Istituto Nanoscienze-CNR and Scuola Normale Superiore, I-56127 Pisa, Italy}
\address{Department of Physics and Astronomy, University of Florence, I-50019, Sesto Fiorentino (FI), Italy}
\affiliation{INFN - Sezione di Pisa, I-56127 Pisa, Italy}
\author{F.~Giazotto}
\affiliation{NEST, Istituto Nanoscienze-CNR and Scuola Normale Superiore, I-56127 Pisa, Italy}

\title{Photonic heat rectification in a coupled qubits system}

\begin{abstract}
We theoretically investigate a quantum heat diode based on two interacting flux qubits coupled to two heat baths. Rectification of heat currents is achieved by asymmetrically coupling the qubits to the reservoirs modelled as dissipative $RLC$ resonators. We find that the coherent interaction between the qubits can be exploited to enhance the rectification factor, which otherwise would be constrained by the baths temperatures and couplings. Remarkably high values of rectification ratio up to $\mathcal R \sim 3.5$ can be obtained for realistic system parameters, with an enhancement up to $\sim 230\%$ compared to the non-interacting case. The system features the possibility of manipulating both the rectification amplitude and direction, allowing for an enhancement or suppression of the heat flow to a chosen bath. For the regime of parameters in which rectification is maximized, we find a significant increase of the rectification above a critical interaction value which corresponds to the onset of a non vanishing entanglement in the system. Finally, we discuss the dependence of the rectification factor on the bath temperatures and couplings.
\end{abstract}
\maketitle

\begin{figure}[t]
\includegraphics{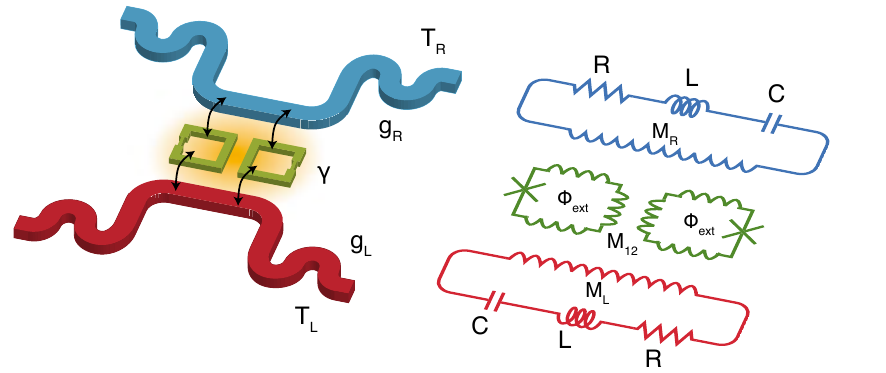}
\caption{Left: graphic illustration of the qubit heat diode. Two interacting flux qubits (green) are mutually coupled to each others via the inductance $M_{12}$ and to two thermal reservoirs with coupling factors $g_{L,R}$. The two reservoirs reside at  temperature $T_L$ and $T_R$ (red and blue). Right: circuit diagram corresponding to the investigated system.}
\label{fig_1}
\end{figure}
\section{Introduction}
The recent development of quantum technologies brought an increasing interest in the experimental and theoretical investigation of heat transport at the nanoscale~\cite{pekola_2015, anders_2017, fornieri_2017, binder_2018}. In this context, phenomena such as phase coherence and entanglement are currently actively studied as they could potentially lead to quantum advantages, e.g., in terms of improved performance of thermal machines~\cite{vischi_2019a} and devices~\cite{martinez-perez_2013, guarcello_2017, hwang_2018, guarcello_2018}, including refrigerators~\cite{niskanen_2007, solinas_2012, brunner_2014,solinas_2016}, heat switches~\cite{ojanen_2008, sothmann_2017, karimi_2017, ronzani_2018, dutta_2020}, heat engines~\cite{hofer_2015, marchegiani_2016, samuelsson_2017, haack_2019, erdman_2019,scharf2020topological,marchegiani2020superconducting}, thermal accelerators~\cite{buffoni_2020}, and towards genuine quantum thermal machines producing entanglement~\cite{brask_2015, khandelwal_2020, aguilar_2020}. A topic of great interest in this context is thermal rectification i.e., the lack of reversal symmetry of heat current under the reversal of the temperature gradient established between two thermal reservoirs. A finite rectification means that the magnitude of the thermal powers changes as the direction of the heat current gets reversed. Superconducting hybrid devices offer outstanding performance regarding electronic heat rectification and highest values of rectification have been reported in systems composed by tunnel junctions between different superconductors/normal metals~\cite{giazotto_2013, martinez-perez_2013a, fornieri_2014, fornieri_2017, martinez-perez_2015}, topological insulators~\cite{bours_2019} and ferromagnetic insulators~\cite{giazotto_2020}. 
While typically electronic heat conduction is considered at low temperature, also the radiative channel can be significant or even dominant in certain designs~\cite{schmidt_2004, meschke_2006, ojanen_2008, ruokola_2009, marchegiani_2021,bosisio2016photonic}. Photonic heat flow is important, for instance, when applying circuit quantum electrodynamics (cQED) schemes to the thermal regime with the potential to study quantum heat transport with remarkable control and precision~\cite{campisi_2013, pekola_2015, pekola_2020}. This emerging field of superconducting circuit quantum thermodynamics (cQTD) has already achieved a number of relevant results~\cite{partanen_2016, ronzani_2018, senior_2020}, being significant for both fundamental study of quantum mechanics as well as for real world quantum technologies applications. The investigation of more complex schemes where the interaction and coherence among multiple qubits plays a prominent role is now actively developing~\cite{campisi_2015, jamshidifarsani_2019, clivaz_2019, khandelwal_2020, tavakoli_2020, rignon-bret_2020}.

Here, we theoretically analyze a prototypical system consisting of two interacting flux qubits coupled to two environmental photonic baths, as sketched in Fig.~\ref{fig_1}. The two qubits are inductively coupled to each other and asymmetrically to the two baths which are modeled as $RLC$ oscillators. This system, which has been previously investigated as a heat switch~\cite{karimi_2017}, can behave as a photonic thermal diode whose rectification factor can be greatly enhanced by the qubits interaction. Moreover, we show that not only the amplitude, but also the direction of rectification can be manipulated, allowing to switch from configurations in which heat flow is favored or suppressed. The high tunability provided by the magnetic flux provides a convenient knob for the control of both the rectifying amplitude and direction. Tunable inductive couplings can allow a further control of the mutual interaction between the qubits themeselves and between qubits and reservoirs~\cite{schwarz_2013}.

\section{Model}
The full Hamiltonian describing the two interacting qubits with the dissipative environments reads
\begin{equation}
H=H_S + H_{S, L}+H_{S, R} + H_L + H_R,
\end{equation}
where $H_S$ is the Hamiltonian of the two interacting qubits, $H_{S,L/R}$ are the qubit-bath interaction and $H_L$ and $H_R$ are the bare baths Hamiltonians, which we shall model as sets of harmonic oscillators.
The Hamiltonian of the interacting qubits reads
\begin{equation}
H_S = H_0+H_{12},
\label{eq:HS}
\end{equation}
with $H_0$ being the two non-interacting flux qubits Hamiltonian and $H_{12}$ the interaction between them. The first term reads~\cite{orlando_1999}
\begin{equation}
H_0 = \sum_{i=1}^2 -\epsilon_i (q\hat\sigma_{z,i} + \Delta_i \hat\sigma_{x,i}),
\end{equation}
where $\epsilon_i=I_{p,i}\Phi_0$, with $I_{p,i}$ the circulating current and $\Phi_0$ the superconducting flux quantum, $\Delta_i$ are the dimensionless tunneling amplitudes, $q=(\Phi_{ex}/\Phi_0-\tfrac{1}{2})$ is the dimensionless applied external magnetic flux and $\hat \sigma_{\alpha, i}$ with $\alpha=\{x,y,z\}$ denote the Pauli matrices of qubit $i$. The qubits are inductively coupled to each other so that the corresponding interaction takes the form
\begin{equation}
H_{12} = \gamma \hat\sigma_{z,1}\hat\sigma_{z,2},
\end{equation}
where $\gamma = 2M_{12}I_{p,1}I_{p,2}$ is the coupling strenght and $M_{12}$ is the mutual inductance between the qubits. In the following, we will consider the condition of identical qubits, i.e., $\epsilon_1=\epsilon_2=\epsilon_0$ and $\Delta_1 =\Delta_2 = \Delta$. As represented schematically in Fig.~\ref{fig_1}, the two qubits are interacting with two distinct heat baths with temperatures $T_L$ (left bath) and $T_R$ (right bath). The dissipative environment can be conveniently modeled as an $LC$ oscillator with a series resistance $R_B$ which inductively couples through the $\hat \sigma_{z,i}$ components of the two qubits as depicted in Fig.~\ref{fig_1}~\cite{storcz_2003, ojanen_2008}. For uncorrelated baths, the Hamiltonian describing the interaction between our two qubit system $S$ and the bath $B=\{L,R\}$ reads~\cite{wilhelm_2003, martinis_2003}
\begin{equation}
H_{S, B} = M_B I_{p}(\hat\sigma_{z,1}+\hat\sigma_{z,2}) \otimes \delta \hat i_{n, B},
\label{eq:bath}
\end{equation}
where the current operator $\delta \hat i_{n,B}$ for the environment $B$ sets the temperature-dependent Johnson-Nyquist noise through its spectral function $S_B$
\begin{subequations}
\begin{equation}\tag{6a}
S_B(\omega) = \int_{-\infty}^\infty dt e^{i\omega (t-t')} \langle \delta \hat i_{n,B}(t) \delta \hat i_{n,B}(t')\rangle.
\end{equation}
The latter can be assessed directly from the impedance of the corresponding environment~\cite{devoret_1997,storcz_2003, ojanen_2008}, which in our case reads
\begin{equation}\tag{6b}
S_B(\omega) = \frac{2\hbar\omega}{1-e^{-\hbar\omega/k_BT_B}} \text{Re}{\{Y_B(\omega)\}},
\end{equation}
\end{subequations}
with $Y_B(\omega)= 1/R_B [1+iQ_B(\omega/\omega_{LC,B}-\omega_{LC,B}/\omega)]^{-1}$ being the admittance of the $RLC$ circuit of resistance $R_B$ and quality factor $Q_B=\sqrt{L_B/C_B}/R_B$. 

\section{Heat transport}
Heat transport is crucially determined, among other quantities, by the rate of transitions incurring in the two-qubit system $S$ as a consequence of their coupling to the noisy reservoirs. By assuming a standard weak-coupling regime, transition rates from level $k$ to level $l$ of the coupled qubits system due to the bath $B$ can be evaluated as~\cite{niskanen_2007, karimi_2017}
\begin{equation}
\Gamma_{k\rightarrow l,B} = \frac{(M_BI_p)^2}{\hbar^2} |\langle k| (\hat\sigma_{z,1}+\hat\sigma_{z,2})	|l\rangle|^2 S_B(\omega_{kl}),
\end{equation}
where $|n\rangle$ denotes an eigenstate of $H_S$, $H_S |n \rangle = E_n |n\rangle$, and $S_B(\omega)$ is the noise spectral function. In order to quantify the thermal power transmitted between the baths, we first need to further evaluate the components of the density matrix $\rho$ of the coupled qubits system. These are governed by the master equation~\cite{breuer_2007, blum_2012, karimi_2017} 
\begin{equation}
\dot \rho_{kl} = -i\omega_{kl}\rho_{kl} + \delta_{kl} \sum_i \rho_{ii} \Gamma_{i\rightarrow k} - \frac{1}{2}\rho_{kl}\sum_i (\Gamma_{k\rightarrow i} + \Gamma_{l\rightarrow i}),
\label{eq:master_eq}
\end{equation}
where $\Gamma_{k\rightarrow l}\equiv\Gamma_{k\rightarrow l,L}+\Gamma_{k\rightarrow l,R}$ are the total transition rates and $\rho$ is written in the instantaneous eigenbasis $|n\rangle$ of the coupled qubits~\footnote{Note that we are describing our system by a global master equation. This is dictated by the geometry of our system, with the two qubits being directly coupled to both baths. For a configuration with the qubits in series rather than in parallel, special care must be taken in choosing whether local or global master equations should be used, see e.g. \citet{hofer_2017, mitchison_2018, cattaneo_2020, khandelwal_2020}.}.
For weak qubit-bath coupling, we can thus write the thermal power to the bath $B$ in the form~\cite{aurell_2019, karimi_2017}
\begin{equation}
P_B = \sum_{i,j} \rho_{ii} E_{ij} \Gamma_{i\rightarrow j,B},
\label{eq_PB}
\end{equation}
where $E_{ij}=E_i-E_j$ is the transition energy between eigenstates $|i\rangle$ and $|j\rangle$ and $\Gamma_{i\rightarrow j, B}$ is the corresponding transition rate induced by the bath $B$.
We can thus quantify the rectification ratio as
\begin{equation}
\mathcal R = \left|\frac{P_B^{+}}{P_B^{-}}\right|,
\label{eq:rect}
\end{equation}
such that the absence of heat rectification corresponds to $\mathcal R = 1$, while $\mathcal R > 1$ or $\mathcal R<1$ indicates a favored or suppressed heat flow to the bath $B$. In the steady state regime, Eq.~(\ref{eq:rect}) can be equivalently expressed in terms of the left/right reservoirs $B = \{L,R\}$.

\begin{figure*}[t]
\includegraphics{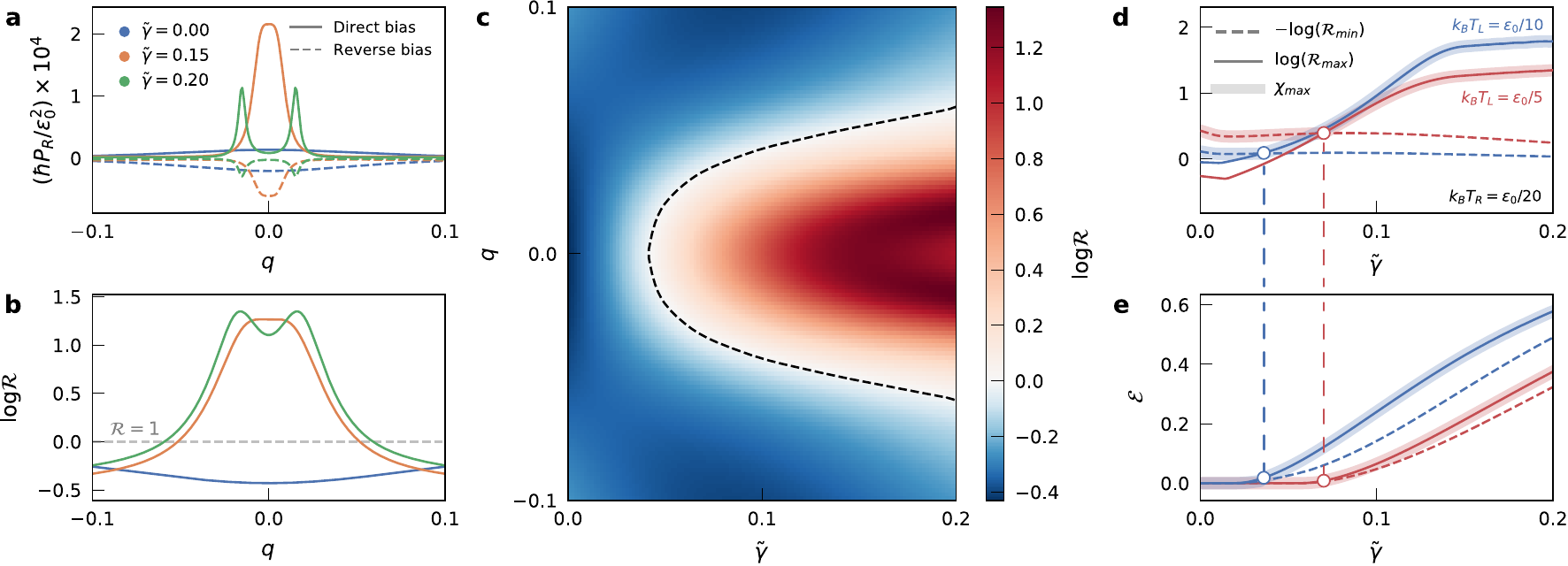}
\caption{a) Normalized power $\hbar P_R^\pm/\epsilon_0^2$ transmitted to the right bath as a function of the dimensionless applied external flux $q$ for different values of interaction $\tilde\gamma=\gamma/\epsilon_0$. Continuous/dashed lines indicates the direct and reverse thermal bias configuration with $k_B T_{L} = \epsilon_0/5$ and $k_B T_{R} = \epsilon_0/20$. b) Rectification ratio $\mathcal R = |P_R^+ / P_R^-|$ extracted from the curves in a). The grey dashed line indicates absence of rectification. c) Full dependence of $\mathcal R$ as a function of $q$ and $\tilde\gamma$. The dashed black line over the white area corresponds to points of absence of rectification ($\mathcal R=1$). Red and blue areas correspond, respectively, to regions of direct and reverse rectification direction.  d) Maximal rectification $\chi_{max}$ as a function of $\tilde\gamma$ for different $T_L$ at fixed $k_B T_{R} = \epsilon_0/20$ is shown as an highlighted curve. The solid and dashed lines corresponds to the quantities $\log{\mathcal R_{max}}$ and $-\log{\mathcal R_{min}}$. The turning points at $\tilde\gamma_c$, associated with a switch of the rectification direction, are shown as white circles. The red curve corresponds to the temperature bias shown in c). e) Entanglement $\mathcal E$ corresponding to the same parameters values of d). In all plots, $\epsilon_0=1$, $\Delta =0.1$, $\hbar \omega_{LC} = 10 \epsilon_0$, $Q_L=Q_R=10$, $R_L=R_R = 1$ $\Omega$, $g_L=0.75$, $g_R=0.25$ are assumed.} 
\label{fig_2}
\end{figure*}
\section{Results}
In the following we shall assume different qubit-bath coupling strengths for the two environments. This provides the necessary structural asymmetry to observe rectification~\cite{segal_2005, ruokola_2009}. For simplicity, we quantify this coupling by the dimensionless parameters
\begin{equation}
g_B = \frac{M_BI_p}{\sqrt{\hbar R_B}},
\end{equation}
which are set to the values $g_R = 0.25$ and $g_L=0.75$. Fig.~\ref{fig_2}a depicts the powers $P_R^\pm$ transmitted to the right bath as a function of the applied dimensionless flux $q$ with direct thermal bias (continuous line), and reverse thermal bias (dashed line). In both cases the dimensionless power  $\hbar P_R^\pm /\epsilon_0^2$ increases dramatically at $\tilde 	\gamma \equiv \gamma/\epsilon_0 = 0.15$ as a result of the resonance condition matched between the qubits energy levels and the frequency of the $LC$-oscillators constituting the dissipative environment~\cite{karimi_2017}. More importantly, a notable variation in the intensity of the power transmitted in the direct/reverse bias configurations is observed and anticipates the significant rectifying properties of our heat diode. In Fig.~\ref{fig_2}b we plot $\log \mathcal R$ as a function of $q$ corresponding to the same parameter values as in Fig.~\ref{fig_2}a. When the qubits are non-interacting (blue line), the heat flow is always favored in the reverse thermal bias configuration, which is testified by the fact that $\log \mathcal R<0$. Moreover, in the simple case of uncoupled qubits, the rectification factor is independent on the number of qubits and depends only on $g_{L,R}$ and $T_{L,R}$~\cite{senior_2020}. Instead, when the qubits are interacting (orange and green lines), a remarkable increase of the rectification factor is observed at $q=0$, with an enhancement for $\mathcal R$ of $\sim 230\%$ at $\tilde\gamma = 0.15$. Moreover, a change in the rectification direction takes place and the system can also be tuned from forward ($\log \mathcal R >0$) to backward ($\log \mathcal R<0$) rectification by spanning $q$. Indeed, it turns out that the non-trivial dependence of the thermal powers on $\tilde\gamma$ can eventually balance and reverse the rectification direction of the qubits with respect to the non-interacting case. This feature allows the system to be fully in-situ tunable both in amplitude and direction of rectification. The complete dependence of $\mathcal R(q,\tilde\gamma)$ is displayed in the colorplot in Fig.~\ref{fig_2}c. The dashed black line over the white area corresponds to a region of absence of rectification, while the blue and red areas correspond to regions of finite rectification, respectively with reverse and direct rectification direction. For the considered system, we clearly observe that one direction is more favorable than the other one (note also the scale on the colorbar). We can further characterize the performance of the heat diode by investigating its points of maximal rectification. In this regard, it is useful to plot the quantities $\mathcal R_{max}(\tilde\gamma) = \max_q \mathcal R(\tilde\gamma)$ and $\mathcal R_{min}(\tilde\gamma) = \min_q \mathcal R(\tilde\gamma)$, which are shown in Fig.~\ref{fig_2}d with continuous and dashed lines for different $T_L$ at fixed $T_R$. The maximal rectification, independently of its direction, is then given by 
\begin{equation}
\chi_{max}(\tilde\gamma) = \max \{ \log \mathcal{R}_{max}(\tilde\gamma),-\log \mathcal{R}_{min}(\tilde\gamma) \}, \end{equation}
which is plotted as an highlighted curve in Fig. 2d.
We observe that for the range of parameters corresponding to $\mathcal R_{min}$, the diode is always rectifying in the reverse thermal bias configuration ($\log \mathcal R_{min} < 0$) and the magnitude of rectification is almost unaffected by $\tilde\gamma$. Differently, $\mathcal R_{max}$ increases in $\tilde\gamma$, starting from a small reverse rectification ($\log \mathcal R_{max} < 0$) at low $\tilde\gamma$ up to a direct rectification ($\log \mathcal R_{max} > 0$) at large $\tilde\gamma$. Interestingly, the turning points $\tilde\gamma_c$ associated with the switch of the rectification direction are followed by a remarkable enhancement of $\mathcal R$, as $\tilde\gamma$ is increased. To characterize the nature of the correlations giving rise to this evolution, we plot in Fig.~\ref{fig_2}e the amount of entanglement between the two qubits at steady state, as quantified by the entanglement of formation $\mathcal E(\rho)$. The latter is standardly defined as $\mathcal E(\rho) = -x \log_2(x) - (1-x)\log_2(1-x) $, where $x=(1+\sqrt{1-C^2})/2$ and $C(\rho)$ is the concurrence associated to the density matrix $\rho$~\cite{wootters_1998}. Remarkably, the critical values $\tilde\gamma_c$ coincide with the values for which the entanglement between the qubits starts to appear, which in turn results to have $\mathcal E = 0.4$ at the higher rectification point shown in Fig.~\ref{fig_2}e (red line). Moreover, despite the reduction of the overall temperature gradient $T_L-T_R$, both $\chi_{max}$ and the amount of entanglement $\mathcal E$ similarly increases for low values of $T_{L}$ as $\tilde\gamma$ increases. For the regime of parameters where $\chi_{max}$ is optimized, the quantum correlations of the system are strong enough to bring the qubits in an entangled state making it possible to envision applications also for quantum information purposes. Although the common features shared between rectification and entanglement veil a deeper connection between the two, an analytical one-to-one correspondence goes beyond the scope of our analysis~\cite{khandelwal_2020}, but might be of interest for further investigations. 

\begin{figure}
\includegraphics{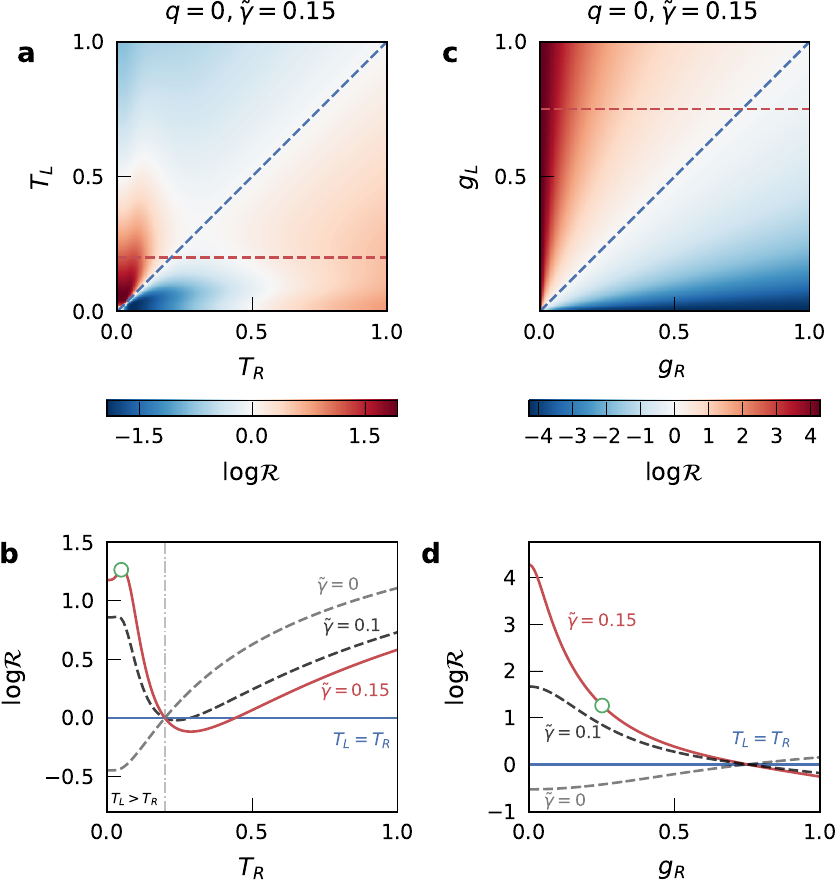}
\caption{a) Dependence of the rectification ratio $\mathcal R$ on the left and right bath temperatures $T_{L,R}$ for $g_L = 0.75$ and $g_R=0.25$. The blue cut corresponds to the axis $T_L=T_R$, while the red one corresponds to fixed left bath temperature $k_BT_L = \epsilon_0/5$. b) The two cuts showed in a) displays the variation of $\mathcal R$ as a function of the right temperature $T_R$. The green circle corresponds to the value $k_B T_R = \epsilon_0/20$ employed in Fig.~\ref{fig_2}. The dashed grey lines corresponds to lower values of interaction $\tilde\gamma=\{0,0.1\}$ at $q=0$. c) Dependence of $\mathcal R$ on the qubit-baths coupling strengths $g_{L,R}$. The blue line corresponds to the symmetric case of identical coupling $g_L=g_R$, while the red line to the condition of fixed $g_L = 0.75$. d) The two cuts displayed in c) showing the dependence of $\mathcal R$ on $g_R$. The green circle corresponds to the condition $g_R=0.25$ presented in Fig.~\ref{fig_2}. The dashed grey lines corresponds to lower interaction values $\tilde\gamma=\{0,0.1\}$ at $q=0$. In all plots, $\epsilon_0=1$, $\Delta =0.1$, $\hbar \omega_{LC} = 10 \epsilon_0$, $Q_L=Q_R=10$, $R_L=R_R = 1$ $\Omega$ are employed.}
\label{fig_3}
\end{figure}
In the following we characterize the performance of the heat diode depending on the temperatures $T_{L,R}$ and the coupling strengths $g_{L,R}$. In particular, Fig.~\ref{fig_3}a shows the dependence of $\log \mathcal R$ on the left and right bath temperatures at the resonance point ($q=0$, $\tilde\gamma=0.15$). The rectification ratio is clearly symmetric with respect to the $T_L = T_R$ axis (blue line), in which no rectification is obviously observed. Higher values of $\mathcal R$ are obtained in the bottom-left angle of the colorplot, corresponding to lower values of baths temperatures as in agreement with the considerations previously done with regard to Fig.~\ref{fig_2}d. The red cut, corresponding to the condition of fixed left bath temperature $T_L$ and variable right bath temperature $T_R$, is plotted in Fig.~\ref{fig_2}b. Notably, the condition $\log \mathcal R=0$, stemming for absence of rectification, is achieved not only in the absence of a temperature gradient $(T_L=T_R)$, but also for another value of temperature bias. For this value, the asymmetry given by the bath temperatures compensates the asymmetric qubits-bath coupling resulting in a overall absence of preferential heat flow. This only holds when the qubits are interacting ($\tilde\gamma \neq 0$) as displayed by the behavior of $\mathcal R$ for $\tilde\gamma=\{0,0.1\}$ with dashed grey lines. Figure~\ref{fig_3}c displays the full dependence of $\mathcal R$ on the qubit-bath coupling strengths $g_{L,R}$ at ($q=0$, $\tilde\gamma = 0.15$). The blue line, corresponding to symmetric coupling $g_L=g_R$, highlights the need for a structural asymmetry between the qubits and baths in order to see any rectification effect~\cite{segal_2005, ruokola_2009}. The behavior of $\mathcal R$ for fixed left bath coupling $g_L=0.75$ is shown in Fig.~\ref{fig_3}c (red line). The rectification is remarkably sensitive to the coupling strengths when the system approaches the resonance condition at $\tilde\gamma=0.15$. In this case, exceptional high values of rectification can be achieved with an enhancement up to $\sim 230\%$ for $g_R=0.25$. Such coupling strengths can be achieved with a proper design of the mutual inductances between qubits and thermal baths. For instance, by assuming a persistent current of $I_p=200$ nA and a bath resistance $R_B=1\Omega$, values of $g_B \sim 0.25 - 0.75$ can be easily achieved for $M_B = 20-45$ pH.

\section{Conclusions}
In conclusion, we have investigated the rectification properties of a system composed of two interacting flux qubits asymmetrically coupled to two $RLC$ resonators acting as thermal baths. The system behaves as an efficient photonic heat diode in which rectification of heat currents between the two thermal environments takes place. We exploit quantum correlations between the two qubits to enhance the rectification factor, which would otherwise be constrained by the coupling to the baths and by their temperatures. Remarkably high values of rectification ratio up to $\mathcal R \sim 3.5$ can be obtained for realistic system parameters, with an enhancement up to $\sim 230\%$ compared to the non-interacting case. The system features the possibility of manipulating both the rectification amplitude and direction, effectively allowing to favor or suppress heat flow to a chosen bath. Standard nanofabrication techniques can be employed for the experimental realization of similar devices. Our analysis can be easily adapted to other kinds of superconducting qubits, different coupling schemes or increased number of qubits. 

\section{Acknowledgments}
We acknowledge the EU’s Horizon 2020 research and innovation program under Grant Agreement No. 800923 (SUPERTED), and the European Research Council under Grant Agreement No. 899315-TERASEC for partial financial support. G.H. acknowledges funding from the Swiss National Science Foundation through the starting grant PRIMA PR00P2 179748 and the NCCR QSIT (Quantum Science and Technology). 

\bibliography{bibliography.bib}
\end{document}